\documentclass{aastex}
\usepackage{ifthen}

\def \version {_working}

\ifthenelse{\equal{\version}{_apj}}
{
\def \figwidth {0.6 \linewidth}
}
{
\usepackage{emulateapj5}
\usepackage{apjfonts}
\usepackage{hyperref}
\def \figwidth {\linewidth}
}
\usepackage{natbib}

\ifthenelse{\equal{\version}{_working}}
{
}
{
\def \today {July 30, 2002}
}

\usepackage{epsfig}
\makeatletter

\newenvironment{inlinefigure}{%
\def\@captype{figure}%
\noindent\begin{minipage}{0.999\linewidth}\begin{center}}
{\end{center}\end{minipage}\smallskip}
\makeatother

\shorttitle{Weak Lensing by X-ray Luminous Clusters -- III.}
\shortauthors{Dahle et al.}
\slugcomment{ApJ, in press}

\usepackage{psfig,epsfig}
\usepackage{amsmath,latexsym}

\begin{document}

\title{Weak Gravitational Lensing by a Sample of \\ X-ray Luminous Clusters of Galaxies -- III. \\ Serendipitous weak lensing detections of dark and luminous mass concentrations.}

\author{H{\aa}kon Dahle\altaffilmark{1,2,3}}
\affil{NORDITA, Blegdamsvej 17, DK-2100, Copenhagen \O, Denmark}
\email{dahle@nordita.dk}
\author{Kristian Pedersen}
\affil{Astronomical Observatory, University of Copenhagen}
\affil{Juliane Maries Vej 30, DK-2100, Copenhagen \O, Denmark} 
\author{Per B. Lilje}
\affil{Institute of Theoretical Astrophysics, University of Oslo}
\affil{P.O. Box 1029, Blindern, N-0315 Oslo, Norway} 
\author{Steve J. Maddox}
\affil{School of Physics and Astronomy, University of Nottingham}
\affil{University Park, Nottingham, NG7 2RD, UK}
\and
\author{Nick Kaiser}
\affil{Institute for Astronomy, University of Hawaii}
\affil{2680 Woodlawn Drive, Honolulu, Hawaii 96822} 

\altaffiltext{1}{Present address: Institute of Theoretical Astrophysics, University of Oslo, P.O. Box 1029, Blindern, N-0315 Oslo, Norway; hdahle@astro.uio.no}
\altaffiltext{2}{Also at: Institute for Astronomy, University of Hawaii}
\altaffiltext{3}{Visiting observer, University of Hawaii 2.24m Telescope at Mauna Kea Observatory, Institute for Astronomy, University of Hawaii}

\begin{abstract} {In the course of a weak gravitational lensing survey of 39 clusters of galaxies,
covering a total sky area of $\sim 1$ square degree,
we have serendipitously discovered mass concentrations
in the fields of \objectname{A1705} and \objectname{A1722} which are most probably not associated
with the main cluster target. By combining weak lensing information with
two-color galaxy photometry in fields centered on our sample clusters,
we identify a new cluster candidate at $z \sim 0.5$ in the field of \objectname{A1705}.
This cluster candidate
also displays strong lensing in the form of a giant luminous arc.
The new mass concentration in the field of \objectname{A1722} also seems to be associated
with an optically luminous cluster of galaxies at $z \sim 0.5$, but in this case there is
some evidence for additional structures along the line of sight that may
contribute to the lensing signal.
A third cluster, \objectname{A959}, has a dark sub-clump which shows
interesting morphological evidence in the mass map for being associated with
the main cluster. This is the first case where there is any significant
evidence for a physical association
between a dark sub-clump (discovered from weak lensing) and a normal cluster.
Analysis of archival X-ray data shows that the three new mass
concentrations are not firmly detected in X-rays and that they are
X-ray underluminous.
}\end{abstract}


\section{Introduction} 
Weak gravitational lensing provides a powerful way to identify cluster-sized
density peaks in the Universe, independent of their baryonic content.
Given the currently modest sky coverage of optical imaging surveys with the
depth and image quality required to detect new clusters by their weak lensing
effect, the number of currently known mass-selected clusters is very small.
It is still an open question whether the mass-selection will lead to the
identification of a population of clusters which are physically
different from optically selected clusters or X-ray detected
clusters. If a population of ``baryon-poor'' clusters is found to exist, they may
be very useful laboratories for the study of dark matter properties.
For instance, some dark matter candidates such as sterile neutrinos may produce
an observable signature from their decay (Abazajian, Fuller, \& Tucker 2001).
The best places
to detect such a signature would be in baryon-poor clusters -- if such objects exist --
where the spectral line corresponding to dark matter decay would be relatively more
conspicuous compared to the emission produced by bremsstrahlung in the hot intra-cluster
gas (Hansen et al.\ 2002).
In any case, the existence of baryon-poor clusters or sub-clusters would pose a
challenge to current models for structure formation.
Furthermore, any previously unrecognized population of clusters with
high mass-to-light ratios would have to be taken into account when using the measured
average mass-to-light ratios of clusters to estimate the density parameter $\Omega_m$.

The present sample of weak lensing-detected clusters is small and contains both
clusters with ``normal'' mass-to-light ratios and objects which appear to be optically dark.
From weak lensing observations in the field of the cluster
\objectname{A1942}, Erben et al.\ (2000) find a secondary mass peak
$\sim 7\arcmin$ south of the cluster center which does not correspond to any
strong concentration of bright galaxies. From deep $H$-band imaging of
this region, Gray et al.\ (2001) constrain the bolometric mass-to-light
ratio to be $M/L_{\rm Bol} > 1000 h$ in solar units for any reasonable lens redshift.
Umetsu \& Futamase (2000) find a dark mass concentration $1\farcm 7$ south
of the high-redshift cluster \objectname{CL1604+4304} ($z=0.90$) with an
estimated mass of $1.2 \times 10^{14} h^{-1} M_{\odot}$ and $M/L_B \geq
1000 h$ in solar units, if it is located at the redshift of \objectname{CL1604+4304}.
At present, it is not clear whether the dark clumps
found by Erben et al.\ and Umetsu \& Futamase are physically associated
with the nearby clusters, or whether they represent chance alignments on the sky of
objects at different redshifts.
Most recently, Miralles et al.\ (2002) have reported evidence for
another dark cluster from a conspicuous alignment of faint galaxies
in a parallel STIS pointing adjacent to the local Seyfert galaxy~\objectname{NGC~625}.

Wittman et al.\ (2001) report the discovery of a more ``normal'' cluster in a ``blank sky'' field through a combination of weak gravitational lensing and photometric data.
Their $BVRI$ photometry shows a concentration of elliptical galaxies close to the lensing-derived mass peak corresponding to the cluster,
and spectroscopic follow-up of candidate cluster members reveals a cluster with modest
galaxy velocity dispersion ($\sigma_v = 615 \pm 150~{\rm km}~{\rm s}^{-1}$) at $z=0.28$.
The mass-to-light ratio of this cluster is $M/L_R = 560\pm 200 h$,
which is somewhat high compared to average values of $M/L_B \approx 300 h$ obtained
from both virial and weak lensing analyses of X-ray selected
clusters (e.g., Carlberg et al.\ 1997; Hoekstra et al.\ 2002; Dahle et al.\ 2003, in prep.), but there are some X-ray selected clusters with similar lensing-derived $M/L$, such as \objectname{MS~1224.7+2007} at
$M/L_R = 640 \pm 150$ (Fahlman et al.\ 1994; Fischer 1999), and
\objectname{A68} and \objectname{A697} at $M/L_R = 680 \pm 230 h$ and
$M/L_R = 450 \pm 115 h$, respectively (Dahle et al.\ 2003, in prep).

In Paper I in this series (Dahle et al.\ 2002) we presented weak lensing measurements
of a sample of 39 X-ray selected clusters. The results were presented in the form of
maps of the reconstructed projected matter density ($\kappa = \Sigma / \Sigma_{\rm crit}$, where $\Sigma_{\rm crit}$ is the critical surface mass density for lensing) and in the form
of radial mass profiles around each (lensing-determined) cluster center.  Several mass maps show
evidence for sub-peaks in the mass distribution which are not
associated with obvious sub-clumping of optically luminous galaxies inside the cluster.
In this paper, we investigate the properties of the most significant of these sub-peaks.
We attempt to constrain the redshift and mass-to-light ratio of these systems and discuss
whether they are likely to be physically associated with their apparent ``host clusters''.

In \S 2 we describe the selection criteria for cluster candidates, which we describe
individually in \S 3. In \S 4 we compare our new (sub-)cluster candidates to those found
by other groups, and compare the observed abundances and physical properties of such
objects to recent theoretical predictions.

The numbers in this paper are given for an Einstein-de Sitter
($\Omega_m = 1$, $\Omega_{\Lambda} = 0$) cosmology.
The Hubble parameter is given by
$H_0 = 100 h\, {\rm km}\, {\rm s}^{-1} {\rm Mpc}^{-1}$, and
all celestial coordinates are given in J2000.0.

\section{Identification of new cluster candidates}
\label{sec:obs}
In Paper I we described a data set of weak lensing measurements of 39 clusters
in the redshift range $0.15 < z < 0.35$, with selection based on very high
X-ray luminosity ($L_{\rm x, 0.1 - 2.4 keV} \geq 10^{45}$ erg s$^{-1}$).
Most of the sky coverage of this survey comes from imaging of
fourteen clusters with the UH8K mosaic CCD camera at the 2.24m University
of Hawaii Telescope at Mauna Kea Observatory, which gives a $19\arcmin$ field when
mounted at the f/10 Cassegrain focus (see Paper I for a complete list of the
observations).
The imaging data were reduced
according to the ``pipeline'' reduction procedure for mosaic CCD data described
by Kaiser et al.\ (1999), and the weak gravitational lensing was measured using
the shear estimator introduced by Kaiser (2000). The reduction and analysis
procedures used are described in detail in Paper I.

In some of the reconstructed mass maps of the fields observed with the UH8K camera,
there are secondary mass peaks which appear significant when compared to peaks in
randomized shear maps (see Figure~48 of paper I), or peaks seen in mass reconstructions
of blank fields (see Wilson, Kaiser \& Luppino 2001),
but do not correspond to concentrations of early-type galaxies at the cluster
redshift. These may be mass concentrations
corresponding to high-redshift clusters that only contribute weakly
to the light distribution in the field, compared to the more nearby
target clusters. At redshifts $> 0.4$, the early-type galaxies at a
given redshift will be significantly bluer than predicted from the
approximate linear $V-I$ color-redshift relation used in Paper I.
Thus, their contribution to the prediction for the dimensionless surface
density given for each cluster in Paper I
will be increasingly underestimated with increasing redshift.

An even more interesting possibility is that these are optically and/or X-ray
dark clusters at
intermediate redshifts (or dark sub-clumps physically associated with their
apparent host clusters), i.e., mass concentrations that have much higher
$M/L$ values than normal clusters.

\subsection{Lensing Cluster Search}
\label{sec:WLsearch}

As noted in Paper I, the $1 \sigma$ level of noise in the
random mass maps is in the range $0.035 < \kappa < 0.05$, mostly depending
on seeing.
Given the large number of fields covered by our survey, we may expect
to see maybe one or two $3\sigma$ peaks that are generated by random noise due
to the intrinsic galaxy shapes.
However, the reality of a mass peak at this level will become
significantly more certain if it can be shown to be strongly correlated with
clustering in the visible galaxy distribution, or with a peak in the
X-ray luminous gas.

From the reconstructions of the dimensionless projected mass surface density
$\kappa$ presented in Paper I, we select candidates for additional mass
concentrations using the following criteria:

\ifthenelse{\equal{\version}{_apj}}
{}{
\begin{figure*}
\centering\epsfig{file=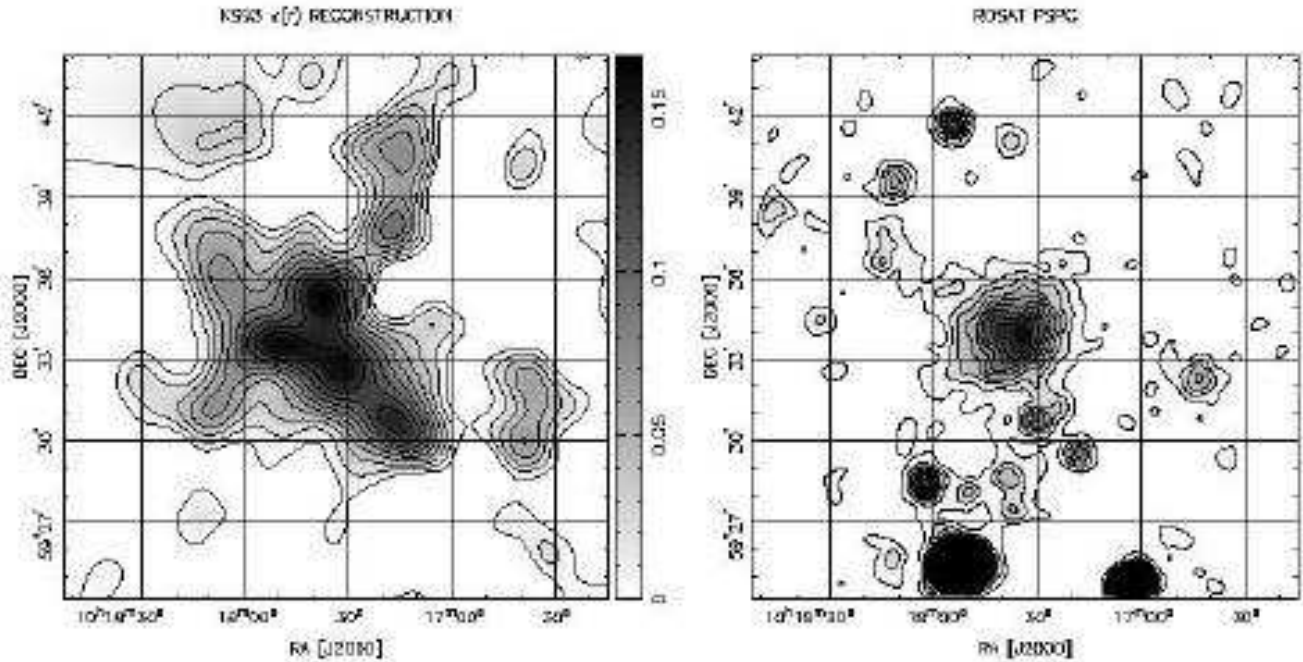,width=\figwidth}
\caption[Mass map of A959.]
{Left panel: The reconstructed projected mass density in the field of \objectname{A959}. The mass map is based on the KS93 inversion method, with a Gaussian smoothing scale of $47\arcsec$.  The scale bar on the right runs from zero to the peak value of $\kappa$. The plotted contours start at zero and are spaced at intervals of $\Delta \kappa = 0.017$. Right panel: A 15.4 ksec ROSAT PSPC exposure in the 0.1 -- 2.4 keV energy band, smoothed with a Gaussian of scale $13\farcs 5$ (9 pixels). The 10 contour levels are plotted on a linear scale ranging from $1.2 \times 10^{-4}$ counts/sec to $9 \times 10^{-4}$ counts/sec.} 
\label{fig:a959_X-ray}
\end{figure*}
}

\begin{itemize}

\item The dimensionless projected surface density $\kappa$ should be
$> 0.15$ (corresponding to a $>3\sigma$ detection of mass
even in the fields with the highest noise) at the position of the peak.
This selection criterion was applied
to mass maps derived separately from $V$- and $I$-band imaging data, and to mass maps
derived from a combination of the two, as detailed in Paper I. We found a total of three
mass peaks which are above the $\kappa$ threshold in at least one of the different mass
maps and which also satisfy the other three selection criteria below.

\item The mass peak should be clearly separated (with offset $> 1\arcmin$) from the
mass peak that corresponds to the light peak/optical
cluster center (or any other significant peak in the light
distribution of galaxies belonging to a previously known cluster, as determined from
their $V-I$ colors).

\item Both $V$- and $I$-band photometry should be available at the position
of the clump. Given the data set of Paper I, this implies that the total sky
coverage of our search is 1.0 deg$^2$. The dominant contribution to the sky coverage
comes from $\sim 19\arcmin$ wide UH8K fields, and all the mass peaks discussed
here are located in such fields.

\item Candidates within $1\arcmin$ from the field edges are discarded, as the
noise in weak lensing reconstructions are known to increase close to
the edges.

\end{itemize}

The limit on the value of the mass surface density $\kappa$ is partly motivated
by the study of Wilson, Kaiser \& Luppino (2001), who present weak lensing data of
similar quality to the data set used in this paper for 1.5 deg$^2$ of ``blank sky''. See
Kaiser, Wilson \& Luppino (2000) and Paper I for more details about the characteristics
of each data set. After smoothing their reconstructed mass maps with a
$45\arcsec$ Gaussian filter
(we use a $47\arcsec$ Gaussian filter) they do not find any pixel values
above $\kappa = 0.15$. If the sky density of optically dark high-density mass peaks
was found to be significantly higher in the fields of known clusters (e.g., the fields studied in
Paper I) than in blank fields (e.g, the fields of Wilson et al.\ 2001), it would be a clear indication 
of a physical association between the dark clumps and ``normal'' clusters.

We note that the mass maps presented in this paper differ slightly from the mass maps
presented for the same clusters in Paper I. The difference is in the relative weighting of
shape measurements of faint galaxies in the two passbands when estimating the shear.
In Paper I, the mass reconstruction was based on shear estimates from a combined ``$V+I$''
catalog which
was made by selecting the catalog entry from the passband in which the object was detected
with the highest significance. The maps presented in this paper are generated by taking
a weighted average (where the weights are given by the ``figure of merit'' values
$\sum Q^2 / d\Omega$; see Paper I and Kaiser 2000) of this ``$V+I$'' mass map and the
mass maps generated separately from the $I$-band and $V$-band catalogs. The $1 \sigma$
level of noise in these maps at the positions of the newly discovered mass peaks is $\kappa = 0.03$.

\subsection{Photometric cluster search}

For the UH8K fields which have both $I$-band and $V$-band data covering
the whole field, the nature of the lensing-derived mass peaks can be investigated by
looking for associated concentrations of red galaxies, corresponding
to galaxy clusters up to $z \sim 1$.

Our photometric procedures are described in Paper I
and are only briefly recounted here. We used SExtractor (Bertin \& Arnouts 1996) to
derive total $V$- and $I$-band magnitudes for galaxies in the fields, and we also
measured a $V-I$ color within a fixed aperture of $2\farcs 7$. The reason for calculating
the color within this smaller aperture rather than based on the total magnitudes is that
the aperture magnitudes are less sensitive to contamination by other objects and to color
gradients within the galaxies themselves and thus yield a tighter red sequence in a
color-magnitude diagram. The magnitudes were corrected for Galactic extinction using
the dust maps of Schlegel, Finkbeiner \& Davis (1998), assuming an $R_V = 3.1$
extinction curve.

To identify optical cluster candidates based on our two-filter photometry,
we used a simplified version of the cluster-red-sequence (CRS) method used by
Gladders \& Yee (2000) for their Red-Sequence Cluster Survey.
This cluster finding algorithm takes advantage of the fact that
early-type galaxies in dense cluster environments form a remarkably tight, almost
horizontal sequence in their color-magnitude diagrams, and these sequences are strongly
homogeneous at a given redshift (e.g., Smail et al.\ 1998; Gladders \& Yee 2000 and
references therein). Also, at redshifts $z \lesssim 1$ the early-type
galaxies that form the red cluster sequence will have a redder $V-I$ color than any
other normal galaxies at redshifts equal to, or lower than, the redshift of the cluster.

\ifthenelse{\equal{\version}{_apj}}
{}{
\begin{figure*}
\centering\epsfig{file=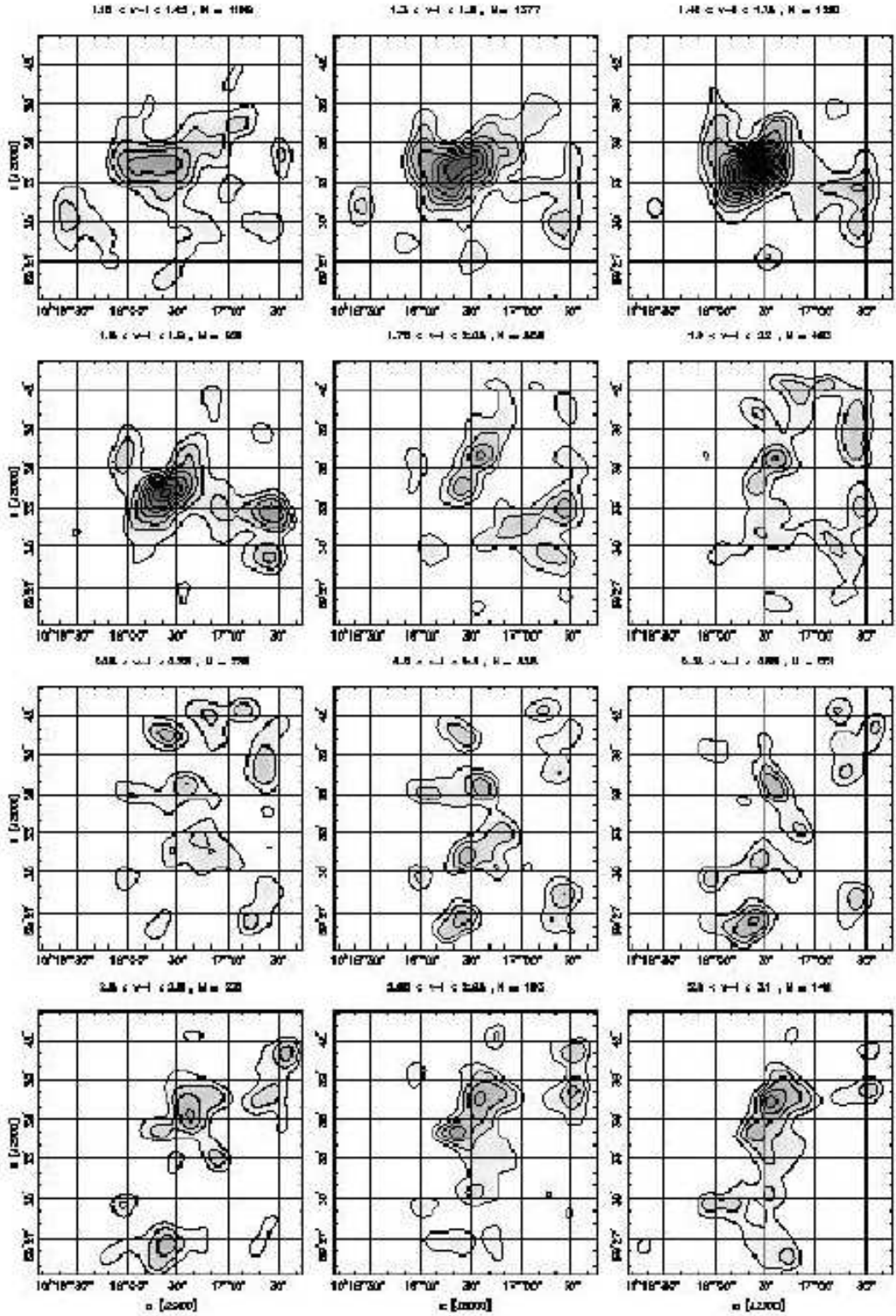,width=15cm}
\caption[Color-selected galaxy density distribution around A959.]
{The surface number density distribution of $I < 23.0$ galaxies in various $V-I$ color intervals in the field of \objectname{A959}. Shown above each plot is the color interval and the number of galaxies in that interval. The plots have been smoothed with a Gaussian of scale $47\arcsec$ to yield a resolution similar to the mass map. The average level has been set to zero, and the contour levels are plotted at 1,2,3,... times the rms fluctuation in field galaxy density in the given color interval. The lack of galaxies in the upper left (northeastern) part of the field is an effect of the missing data from chip 4 in the UH8K mosaic.}
\label{fig:a959_colnumdens}
\end{figure*}
}
The basic idea behind the CRS method is to create a series of smoothed maps of the galaxy
density in the field by selecting galaxies in a series of partially overlapping slices
in the color-magnitude diagram.  Gladders \& Yee (2000) used color slices defined by
theoretical predictions of the red sequence, but we choose here to neglect the slope of
the red sequence and use a series of horizontal slices evenly spaced in $V-I$. Both
theoretical predictions from population synthesis models and empirical results based
on our data set show the absolute value of the red-sequence slope to be less than 0.1
per $I$-magnitude. Given the relatively narrow magnitude range (particularly at high
redshifts) of cluster galaxies contributing to a given slice, and the fact that our color
slices are approximately twice as wide as those used by Gladders \& Yee (2000), the effect
of the slope will not significantly influence our ability to detect cluster candidates with
this algorithm. However, it would probably bias redshift
estimates of cluster candidates, but only at levels comparable to the systematic
uncertainty introduced by the photometric calibration of the UH8K mosaic camera (we note
that the photometric stability of the CCD chips in the UH8K is questionable at the
$\lesssim 0.1$ magnitude level).

For each slice, we include all galaxies at $I<23$ that have a probability higher than 10\%
of really belonging to that slice when taking the photometric uncertainties into account.
The width of the slices was $\Delta (V-I) = 0.3$ at bright magnitudes and slightly larger
at faint magnitudes due to the increasing photometric uncertainties. This is about twice
the measured intrinsic width of the red sequence (Stanford, Eisenhardt, \& Dickinson 1998).

Rather than using a more elaborate galaxy weighting scheme such as the one outlined by
Gladders \& Yee (2000), we chose a simple faint galaxy cutoff at $I = 23$. Also, to
generate smoothed galaxy number density plots comparable to the weak lensing-derived mass maps,
we employed a Gaussian smoothing kernel of fixed scale $47\arcsec$ rather than a kernel
tailored to fit the expected cluster density profile at a given redshift.
The primary purpose of generating these plots is to search for galaxy density enhancements
at the position of the lensing-detected cluster candidates.
 
\subsection{Lensing masses and mass-to-light ratios}

For each (sub-)cluster candidate, the (minimum) mass and the mass-to-light ratio were
estimated within an aperture centered on the corresponding peak found in the
$\kappa$ map (see Table~\ref{tab:clumpdata}).
A lower limit on the aperture mass was found using the aperture densitometry
statistic $\zeta$ of Kaiser et al.\ (1994):

\begin{equation}
\zeta (R_1 , R_2) = \bar{\kappa} (R_1) - \bar{\kappa} (R_1 , R_2) = \frac{2}{1 - R^2_1/R^2_2} \int_{R_1}^{R_2} \langle \gamma_T \rangle d \ln r ,
\label{eq:apdens2}
\end{equation}

\noindent
which measures the average surface mass density within an
aperture radius $R_{\rm ap} = R_1$ {\it minus} the average surface mass density in a
surrounding annulus $R_1 < R < R_2$ (here, $\gamma_T$ is the tangential component of the
shear). Hence, a lower limit on the aperture mass can be simply calculated as
$M_{\rm ap} = \pi r^{2} \zeta \Sigma_{\rm crit}$. We chose $R_2 = 2 R_1$, and our
estimate of $M_{\rm ap}$ will in this case e.g., underestimate the true aperture mass by
33\% for an isothermal sphere ($\Sigma \propto R^{-1}$) mass distribution.

In order to minimize the bias caused by the presence of a more
massive cluster in each field we chose to use an outer radius $R_2$ much smaller
than the $R_2$ values used for the aperture mass measurements presented in Paper I.
For this reason, $R_2$ was chosen to be less than half the angular separation between
the new cluster candidate and the Abell cluster. For this geometry, we find from
simple simulations that the aperture mass measurement is affected by
$ < 1 $\% , provided that the mass of the Abell cluster is not more than twice that of
the new cluster candidate.
In the case of the \objectname{A959} and \objectname{A1705} fields, we used a control
annulus with $R_2 = 2\arcmin$ and inner radius $R_1 = 1\arcmin$. The cluster candidate in
the \objectname{A1722} field had a somewhat larger separation from its host cluster,
and it was therefore possible to use a larger aperture with $R_2 = 4\arcmin$ and
$R_1 = 2\arcmin$. By defining an aperture measurement of the surface light density
$\zeta_L$ in a manner similar to equation~(\ref{eq:apdens2})
(i.e., generating an aperture luminosity estimate by
subtracting the average galaxy luminosity density in the annulus $R_1 < R < R_2$ from the
average galaxy luminosity density at $R < R_1$), the mass-to-light ratio of the mass peak
can be estimated. Thus, in contrast to $M_{\rm ap}$, this ratio will not be
systematically underestimated.

\subsection{Hot gas content}
In X-ray luminous clusters of galaxies, such as those in the cluster
sample of Paper I, most of the baryonic mass is in the form of hot,
X-ray emitting gas (e.g., Ettori \& Fabian 1999).
In order to constrain the hot gas content of the (sub-)cluster
candidates we analyzed the best quality archive X-ray data
(standard screened EINSTEIN IPC, ROSAT PSPC and ASCA SIS data,
respectively) for these candidates. No obvious X-ray emission is
associated with any of the (sub-)cluster candidates so we determine
upper limits for their X-ray luminosity as described below.

An upper limit for the (sub-)cluster X-ray count rate
is extracted within two apertures: The aperture as determined
from the weak lensing maps ($R_{\rm ap}$) and the radius within
which the average (sub-)cluster mass density is 500 times the critical density
of the Universe ($r_{500}$). Special care is taken in estimating the background.

For EINSTEIN IPC data (A1705) and ROSAT PSPC data (A959) the
background is estimated from the data set including the
new mass concentration itself. The background is calculated as the average
of three source-free regions of the same size as the (sub-)cluster region,
and at the same optical axis distance.

For ASCA SIS data (A1722) the background count rate is evaluated
from the publically available ``blank sky'' data in a
two stage process: A first background estimate is obtained using exactly the
same region on the SIS chip as the (sub-)cluster region.
Then, the general background level in the ``blank sky'' data is compared to
the general background level in the (sub-)cluster data by calculating the
average count rate within three regions of the same size as the region
of interest, at the same optical axis distance and devoid of obvious
X-ray sources.
The inital background estimate is then corrected for the average difference
in background level between the ``blank sky'' data and the
(sub-)cluster data. Finally, data from the SIS0 detector and the SIS1
detector are merged into one common SIS data set.

For each (sub-)cluster, we derive net count rates (corrected for point source
contributions) within the two apertures
and convert these to 0.1 -- 2.4 keV fluxes using PIMMS
and an absorbed Raymond-Smith spectral model with the following parameters:
Abundance is $Z=0.25 Z_{\sun}$, absorption by neutral hydrogen is fixed
at the Galactic value (from Dickey and Lockman 1990), and for each
(sub-)cluster
the temperature is fixed at the value derived from the weak lensing mass within
$r_{500}$ using the mass-temperature relation of Finoguenov et al.\ (2001).

Assuming the redshift of the (sub-)cluster given in Table~\ref{tab:clumpdata}
based on optical data, the flux is converted to a luminosity
for the desired cosmology.
In order to determine the X-ray luminosity of the (sub-)clusters relative
to a sample of nearby ``standard'' clusters, the upper limit on the 0.1 -- 2.4 keV
luminosity within $r_{500}$ was compared to the expected value based on
the lensing mass within $r_{500}$ and the mass-luminosity relation of
Reiprich \& B{\"o}hringer (2002) (best bisector fit).

\section{Candidate Mass Concentrations} 

The selection criteria outlined in \S~\ref{sec:WLsearch} produced three new cluster candidates,
in the fields of \objectname{A959}, \objectname{A1705} and \objectname{A1722}. To make the
following discussion clearer and more concise, we introduce the naming convention
``WL HHMM.M(+/-)DDMM'' for these objects, where ``WL'' denotes a weak lensing-detected mass concentration
and is followed by numbers that are based on the celestial coordinates of the objects (for epoch J2000.0),
where the right ascension is given in units of hours and minutes (with one decimal) and the
declination is given in units of degrees and arcminutes.
For each candidate, we analyze its optical and lensing properties based on the data
presented in Paper I along with X-ray data to search for emission from any
associated hot gas. Table~\ref{tab:clumpdata} summarizes the most important parameters for each cluster candidate.

\ifthenelse{\equal{\version}{_apj}}
{}{
\begin{table*}[htbp!]
\begin{center}
\caption{Properties of the weak lensing-detected mass concentrations}
\begin{tabular}{cccccccccc}
\hline
\hline
Name & $\alpha$ & $\delta$ & Redshift & $R_{\rm ap}$ & $M_{\rm ap}$ & $M/L_B$ & $\sigma_{\rm DM}$ & $L_{\rm X,ap}$ \\
  & (J2000.0) & (J2000.0) & $z_{est}$ & ($h^{-1}$ kpc) & ($10^{14}h^{-1} M_{\sun}$) & ($h [M/L_B]_{\sun}$) & (km s$^{-1}$) & ($h^{-2}$ $10^{44} {\rm erg~s}^{-1}$) \\
\hline
 \objectname{WL 1017.3+5931} & 10 17 15 & 59 30 46 & 0.286? & 161 &
$1.1\pm 0.3$ & $> 500$ & $780^{+120}_{-160}$ & $<0.008 \pm 0.006$ \cr
 \objectname{WL 1312.5+7252}  & 13 12 31 & 72 51 32 & 0.55 & 222 &
$2.7\pm 0.8$ & $173\pm 54$ & $1150^{+160}_{-190}$& $<0.09 \pm 0.03$\cr
 \objectname{WL 1320.4+6959}  & 13 20 22 & 69 58 43 & 0.45 & 409 &
$4.8\pm 1.9$ & $296\pm 118$ & $940^{+160}_{-210}$ & $<0.20 \pm 0.1$ \cr
\hline
\end{tabular}
\tablecomments{Units of right ascension are hours, minutes and seconds, and units of declination are degrees, arcminutes, and arcseconds. The X-ray luminosity is given in the 0.1-2.4~keV band, and the uncertainty arises from poisson noise as well as point source subtraction. The values are given for an Einstein-de Sitter Universe. The source redshifts adopted for the $M_{\rm ap}$ estimates are based on photometric redshift data from the Hubble Deep Field (see Eq.~17 of Paper I), and correspond to having a single-screen source population at an effective redshift of $z_s = 0.7$, $z_s = 0.9$, and $z_s = 0.8$ for \objectname{WL 1017.3+5931}, \objectname{WL 1312.5+7252}, and \objectname{WL 1320.4+6959}, respectively}
\label{tab:clumpdata}
\end{center}
\end{table*}
}

\subsection{The A959 Field}
\label{sec:a959field}

From spectroscopic observations at the WHT,
we have determined a redshift $z\!=\!0.286$ for this cluster
(Irgens et al.\ 2002), significantly lower than the value
usually quoted in the literature ($z = 0.353$). After properly revising
its X-ray luminosity and temperature, it is found to have
$L_{\rm x, 0.1-2.4 keV} = 1.43 \times 10^{45}$ erg s$^{-1}$
(B{\"o}hringer et al.\ 2000) and $T = 6.6$~keV (Mushotzsky \& Scharf 1997).
A singular isothermal sphere (SIS)-model fit to the tangential shear measured
out to a radius of $\sim 1.5 h^{-1}$Mpc shows that the cluster has
a mass corresponding to a velocity dispersion
$\sigma = 990^{+100}_{-110}$km~s$^{-1}$.
The fit to a SIS-model is however not very good, and weak lensing mass
measurements at large radii indicate that this cluster is very massive, with a
projected 2D aperture mass $M_{\rm ap}(< 1.5 h^{-1} {\rm Mpc}) \sim 2 \times 10^{15} M_{\odot}$ (see Paper I).

\ifthenelse{\equal{\version}{_apj}}
{}{
\begin{inlinefigure}
\centering\epsfig{file=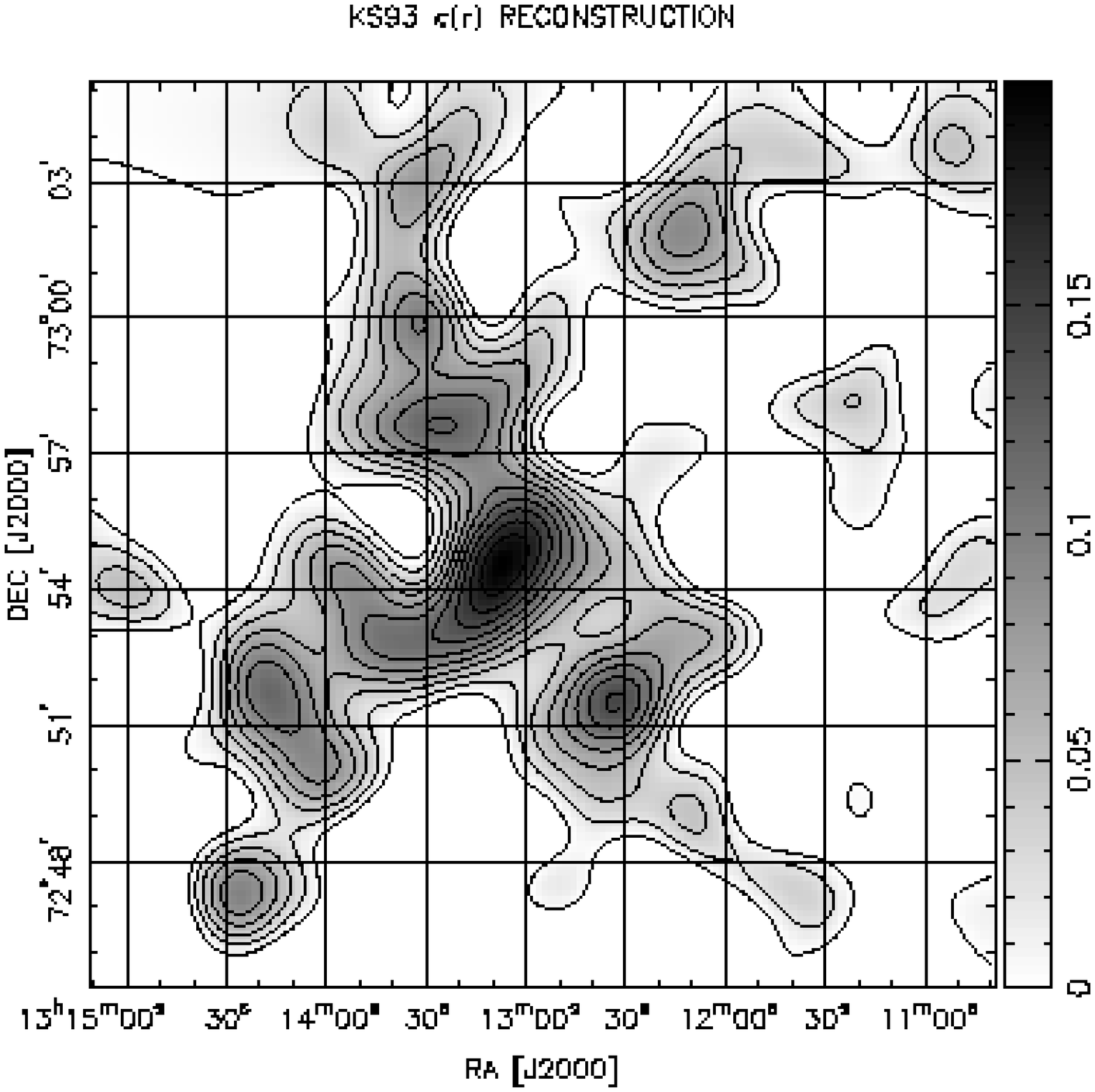,width=\figwidth}
\caption[Mass map of A1705.]
{The reconstructed projected mass density in the field of \objectname{A1705}. The mass map is based on the KS93 inversion method, with a Gaussian smoothing scale of $47\arcsec$.  The scale bar on the right runs from zero to the peak value of $\kappa$. The plotted contours start at zero and are spaced at intervals of $\Delta \kappa = 0.020$.}
\label{fig:a1705_massmap}
\end{inlinefigure}
}

\ifthenelse{\equal{\version}{_apj}}
{}{
\begin{figure*}
\centering\epsfig{file=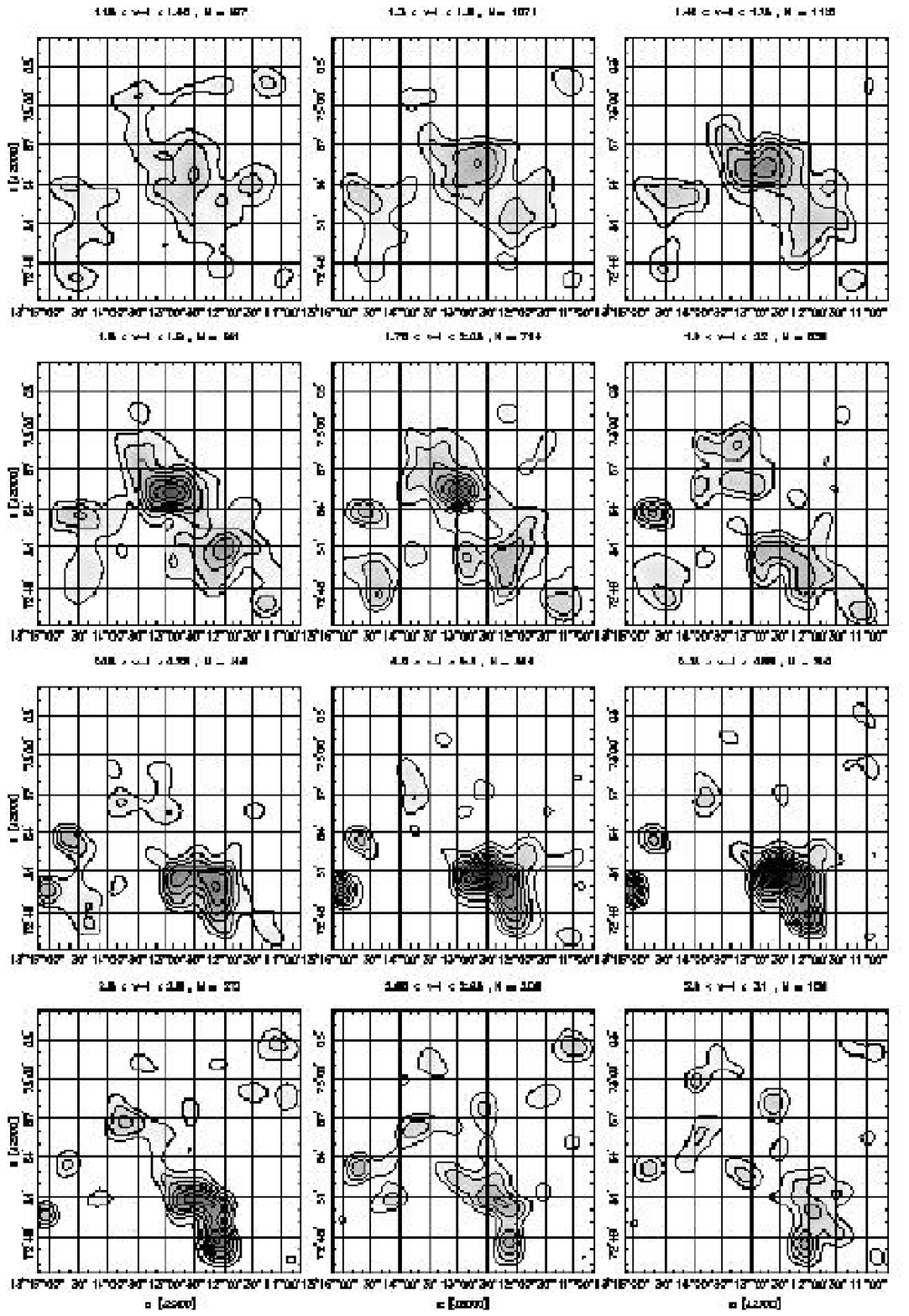,width=15cm}
\caption[Color-selected galaxy density distribution around A1705.]
{The surface number density distribution of $I < 23.0$ galaxies in various $V-I$ color intervals in the field of \objectname{A1705}. See the caption of Fig.~\ref{fig:a959_colnumdens} for details.}
\label{fig:a1705_colnumdens}
\end{figure*}
}

Optically, the cluster center is not dominated by any single galaxy,
but it has a core region consisting of many early-type galaxies of
similar brightnesses.  A highly significant mass peak (with some evidence
for substructure) is seen in the weak lensing mass map
(see Figure~\ref{fig:a959_X-ray}), and the dark matter distribution
appears to resemble the cluster light distribution,
as apparent in the first four panels of Figure~\ref{fig:a959_colnumdens}.
However, there is a dark matter concentration, hereafter denoted as \objectname{WL 1017.3+5931},
$\sim 6\arcmin$ southwest of the cluster center which does not
correspond to any peak in the luminosity distribution. Its peak value\footnote{%
As noted in \S~\ref{sec:WLsearch}, the mass maps used for the selection of mass peaks
were different from those displayed in this paper. The reason for not using the new maps presented here
(and lowering the selection threshold from $\kappa=0.15$ to $0.14$, which would have selected
exactly the same three objects!) is that the noise is more clearly defined
in the old maps. While the new maps have lower noise, this is not achieved
by increasing the available sample of background galaxies, but instead by in
effect taking a weighted average of the shape measurements of these galaxies in
the two passbands. Hence, the photon counting noise for the shape measurements of
faint galaxies is somewhat reduced, while the noise due to the finite number density of
background galaxies does not change.}
 is $\kappa = 0.143$.

Due to the very low background, the ROSAT PSPC 15.4 ksec exposure yields the tightest current
constraints on diffuse X-ray emission from \objectname{WL 1017.3+5931}. In
Figure~\ref{fig:a959_X-ray}, a couple of X-ray sources are seen
close to \objectname{WL 1017.3+5931}. Archival ROSAT HRI images show that
sources ``C'' and ``D'' are extended sources while source ``A'' is a point source
and source ``B'' consists of two point sources on top of some extended
emission. This extended X-ray emission is possibly linking \objectname{WL 1017.3+5931}
to the main cluster of \objectname{A959}. The mass morphology of
\objectname{WL 1017.3+5931} provides clearer evidence for this link,
since there appears to be a bridge of matter extending from the
\objectname{A959} cluster center toward
\objectname{WL 1017.3+5931}, and the peak itself appears highly elongated in
the direction
toward the center of \objectname{A959}. Even at the lowest ``saddle point'' in
the apparent mass bridge between \objectname{A959} and
\objectname{WL 1017.3+5931}, the projected mass density is still as high as
$\kappa = 0.122$, which is $85\%$ of the peak value at the location of
\objectname{WL 1017.3+5931}. Simulations of the region between two cluster
mass peaks with a Gaussian smoothing scale equal to that of the real mass map
and with random realizations of realistic noise indicate that this feature is
(marginally) significant, at almost the $2\sigma$ level.
Although the reality of this feature is currently debatable, the apparent mass
bridge and the ellipticity of
\objectname{WL 1017.3+5931} probably constitute the only significant evidence
so far for a physical link between a dark sub-clump found by weak lensing in a
cluster field and its apparent host cluster.

No strong galaxy overdensities
are seen at the position of \objectname{WL 1017.3+5931} for any of the color slices plotted in
Figure~\ref{fig:a959_colnumdens}. The lower limit value for $M/L_B$ in
Table~\ref{tab:clumpdata} is based on an aperture light estimate using galaxies
in the $2.0 < V-I < 4.0$ color range. These are all redder than the
\objectname{A959} cluster galaxies (the red cluster sequence has $V-I \simeq 1.7$).
If we instead consider galaxies in the color range of the red cluster sequence of
\objectname{A959}, we find that the luminosity density in the control annulus at
$1\arcmin < R < 2\arcmin$ is higher than the luminosity density within $1\arcmin$, implying that
there is no overdensity of light associated with \objectname{A959} cluster galaxies
at the location of \objectname{WL 1017.3+5931}. Two of the brightest galaxies
in the field ($I \simeq 16.5$) are visible $2\arcmin$ from the center of \objectname{WL 1017.3+5931},
but these are likely to be foreground ellipticals ($V-I \simeq 1.3$), and no significant galaxy
clustering is seen around these two galaxies.
The estimated velocity dispersion of \objectname{WL 1017.3+5931} listed in Table~\ref{tab:clumpdata}
comes from a SIS-model fit to the tangential weak shear averaged in radial bins centered
on the dark clump.
If \objectname{WL 1017.3+5931} is at the \objectname{A959} main cluster
redshift, within a radius of $r_{500}$ it is a
factor three X-ray under-luminous relative to the expectation from the
mass-X-ray luminosity relation of Reiprich and B\"{o}hringer (2002).
However, given the scatter in the mass-luminosity relation of an order of
magnitude, even though \objectname{WL 1017.3+5931} is relatively X-ray dark
it is still consistent with the relation.
Chandra or XMM-Newton observations are required to pin down the
X-ray emission of \objectname{WL 1017.3+5931} and accurately constrain
the hot gas content, ie. if it is really X-ray dark.

In general, there is very good morphological agreement between the
mass map and hot gas as mapped by ROSAT PSPC: The mass extension towards the
northeast has a clear X-ray counterpart, and at the position of the
extended X-ray source ``D'' a mass concentration is seen. This feature
also corresponds to an overdensity of galaxies (most prominent in the $1.6 < V-I < 1.9$ color slice in Figure~\ref{fig:a959_colnumdens}) with colors similar to, or slightly redder than, the early-type galaxies in the core of \objectname{A959}. The general southeast-northwest elongation of the \objectname{A959} cluster galaxy distribution is matched in X-rays.

We also note that there are several concentrations of red ($2.0 < V-I < 3.0$)
galaxies in the field, most clearly visible in the lower three panels of
Figure~\ref{fig:a959_colnumdens}. One of these is only $\sim 1\arcmin$ north
of the \objectname{A959} cluster center (see the $2.65 < V-I < 2.95$ color slice in Figure~\ref{fig:a959_colnumdens}) and may contribute to the substructure seen in the mass peak of \objectname{A959}.

\subsection{The A1705 Field}
\label{sec:a1705field}

Optical images show that the cluster center of \objectname{A1705} is dominated by a single
cD galaxy. The $\kappa $ reconstruction of this field (see Figure~\ref{fig:a1705_massmap})
shows a number of secondary peaks in addition to the most significant
mass peak, which is associated with the center of \objectname{A1705}.
The position given for \objectname{A1705} by Abell (1958) is $5\arcmin$ off from
what we take to be the cluster center, based on the peaks in the mass, light
and galaxy number density distributions, while the Zwicky cluster
\objectname{Zw~1312.1+7311} is more consistent with the position we determine for the
\objectname{A1705} cluster center (at $\alpha = 13^{\rm{h}} 13^{\rm{m}}.0, \delta = 72\arcdeg
55\arcmin$). The color slicing in Figure~\ref{fig:a1705_colnumdens} shows several interesting
effects: Firstly, the morphology-density relation manifests itself in the effect that
the density peak associated with \objectname{A1705} appears more concentrated
as one moves to progressively redder color intervals (which contain the most evolved
early-type cluster galaxies). This effect is also apparent in the similar plots
for \objectname{A959} and \objectname{A1722}. Secondly, the density peak of \objectname{A1705}
in Figure~\ref{fig:a1705_colnumdens} shows a significant shift in the centroid position
as a function of color. The galaxy number density peak is seen to move closer to the position of the
mass peak for redder color intervals, and this again demonstrates that the galaxies with
the most evolved stellar populations are concentrated in the densest environments.
The centroid shift may indicate that \objectname{A1705} is dynamically young.
A secondary mass peak situated about $4\arcmin$ toward the northeast from the main peak
appears to be associated with galaxies
at the cluster redshift (the first five panels of Figure~\ref{fig:a1705_colnumdens}
appear to show an extension toward northeast from the main galaxy
overdensity peak associated with \objectname{A1705}). Close to this secondary
mass peak, and possibly contributing to it, is the $z=0.112$ IRAS galaxy
\objectname{F13121+7315} (at $\alpha = 13^{\rm{h}}
13^{\rm{m}} 32.0^{\rm{s}},\delta = 72\arcdeg 59\arcmin 11\arcsec $),
associated with the AGN X-ray source \objectname{RX~J1313.5+7259}.

\ifthenelse{\equal{\version}{_apj}}
{}{
\begin{inlinefigure}
\centering\epsfig{file=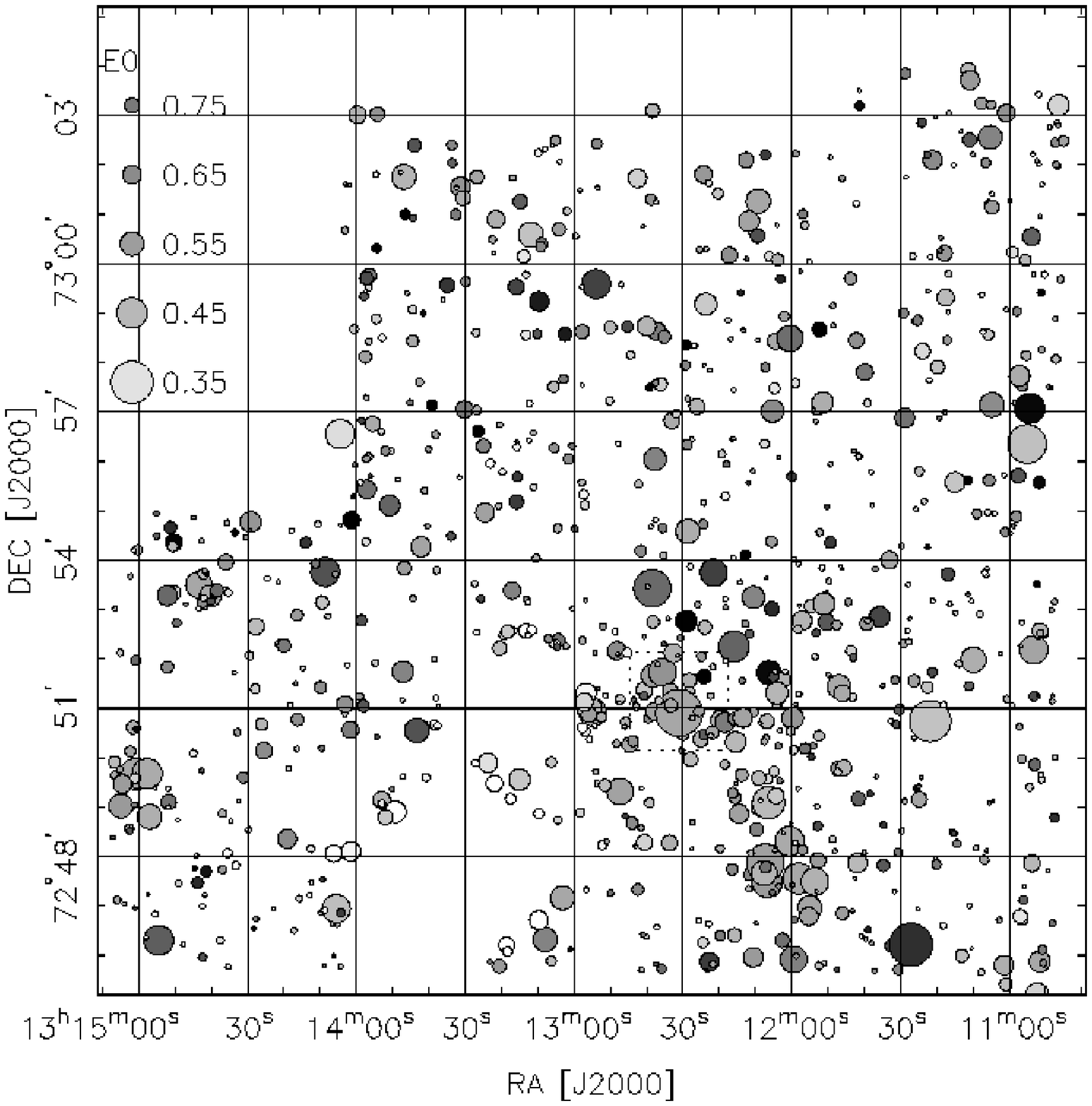,width=\figwidth}
\caption[Red galaxies in the field of A1705.]
{The reddest galaxies ($V-I > 2.2$) 
in the field of Abell 1705. The area of the circles is proportional to
the $I$-band flux. The legend in the upper left corner indicates the
flux and color (assuming no evolution of the SED) of an $L_{\ast}$ E0 galaxy at a given redshift. The apparent lack of galaxies around $\alpha = 13^{\rm{h}} 13^{\rm{m}} 30^{\rm{s}},\delta = 72\arcdeg 48\arcmin$ is an artifact caused by a $V=7$ star at this position. The square region denoted by the dotted lines indicates the region shown in Fig.~\ref{fig:a1705_newcluster_image}.}
\label{fig:a1705_colsel}
\end{inlinefigure}
}

The two lowest rows of the galaxy number density plots in
Figure~\ref{fig:a1705_colnumdens} shows a strong peak southwest
of the location of the main cluster center. This is a galaxy concentration
at a redshift well beyond \objectname{A1705}.
Figure~\ref{fig:a1705_colsel} also shows this strong overdensity in the
distribution of the reddest ($V-I > 2.2$) galaxies in the field. A strong
concentration of galaxies with similar colors is centered on
what appears to be a distant giant elliptical galaxy located
at $\alpha = 13^{\rm{h}} 12^{\rm{m}} 31.1^{\rm{s}},\delta = 72\arcdeg
50\arcmin 54\arcsec$. A peak in the mass distribution, which we have named
\objectname{WL 1312.5+7252}, is situated less than $1\arcmin$ away from this galaxy.
Its peak value is $\kappa = 0.143$. A rough cluster redshift estimate can be
derived for this peak from the median color ($V-I = 2.45$) of the early-type cluster
galaxy sequence seen in a color-magnitude diagram.
Using our empirically calibrated color-redshift relation for
early-type cluster galaxies (see Figure~47 of Paper I), we estimate the
redshift of the cluster associated with \objectname{WL 1312.5+7252} to be $z=0.55\pm 0.05$.
This may be a slight underestimate of the
true redshift, since cluster ellipticals at $z \gtrsim 0.5$
are less evolved and intrinsically slightly bluer than their counterparts
at lower redshifts (Lubin 1996).

A curved blue object, apparently
a gravitationally lensed arc, is situated at $\alpha = 13^{\rm{h}} 12^{\rm{m}}
32.6^{\rm{s}},\delta = 72\arcdeg 50\arcmin 55\arcsec$, $7\arcsec$ east of the
distant giant elliptical galaxy (see Figure~\ref{fig:a1705_newcluster_image}).
This arc is quite bright (at $I = 20.90$, $V = 21.39$), and
it is worth noting that the comparatively bright arc magnitude places
 \objectname{WL 1312.5+7252} among the very small number of clusters at $z>0.5$ that
are known
to produce strongly lensed arcs that are sufficiently bright for spectroscopic studies.
The only other such examples known to us are \objectname{CL2236-04}
(Melnick et al.\ 1993) and \objectname{RCS 0224-0002} (Gladders, Yee, \& Ellingson 2002).

If this arc is indeed a strongly lensed galaxy, the separation between the
arc and the giant elliptical galaxy can be used to make a rough estimate of the mass of
 \objectname{WL 1312.5+7252}.

\ifthenelse{\equal{\version}{_apj}}
{}{
\begin{inlinefigure}
\centering\epsfig{file=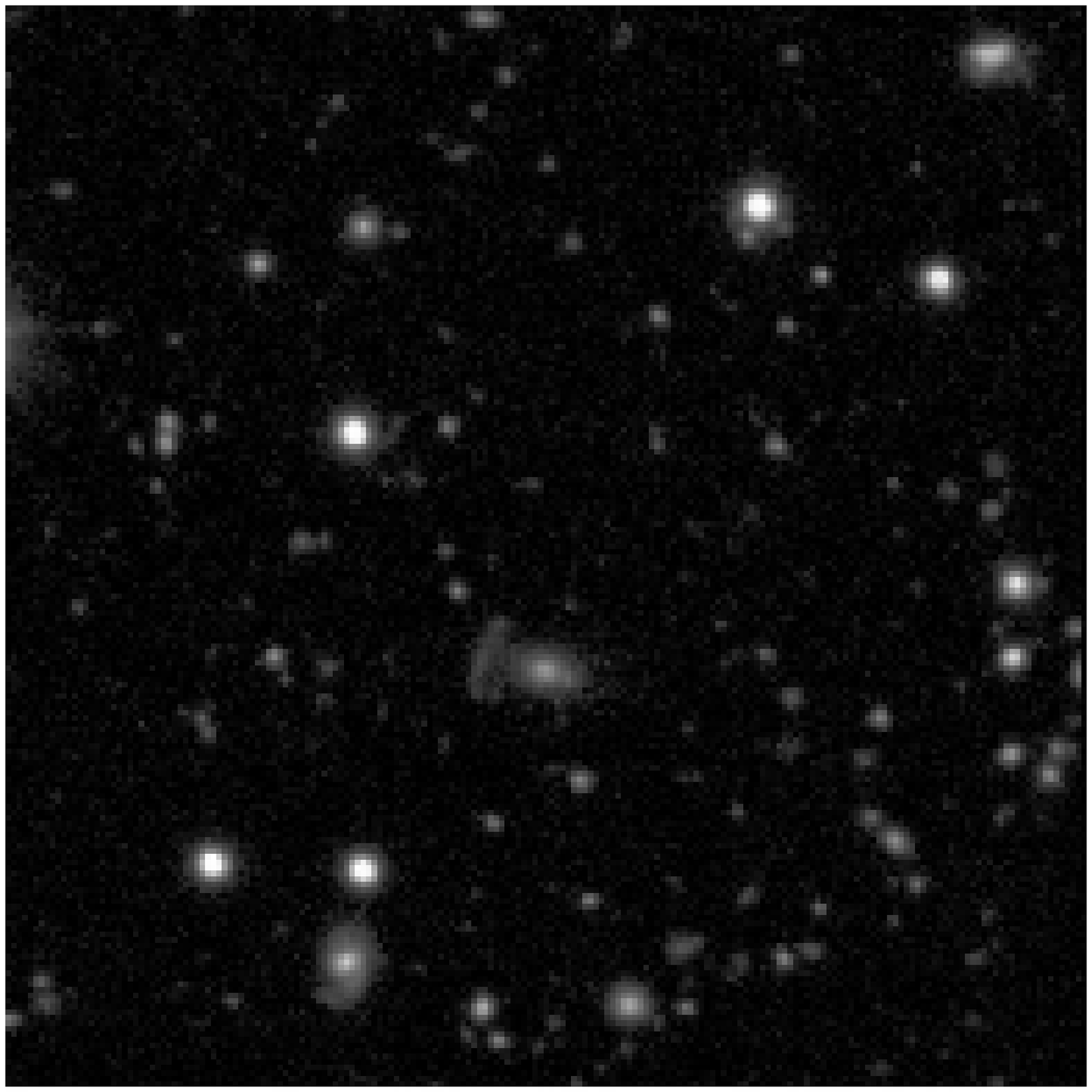,width=\figwidth}
\caption[Center of candidate cluster with gravitational arc.]
{A blue arc curving around the central galaxy of the $z = 0.55$ cluster 
candidate \objectname{WL 1312.5+7252} in the field of Abell 1705. 
North is up; east to the left and the FOV is $2\arcmin \times 2\arcmin$. This area is marked by the dotted 
lines in Fig.~\ref{fig:a1705_colsel}.}
\label{fig:a1705_newcluster_image}
\end{inlinefigure}
}

Assuming a SIS-type mass distribution centered on
a giant elliptical galaxy at redshift $z_l = 0.6$, and assuming that the
lensed arc is situated at the Einstein radius $\theta_{E}$, we estimate a
velocity dispersion $\sigma_{\rm arc} = c \sqrt{(\theta_{E} w_s)/(4\pi w_{ls})}
  \simeq 900 $km s$^{-1}$ for $z_{s} =$ 1 and $\sigma_{\rm arc} \simeq 700 $km s$^{-1}$
for $z_{s} =$ 2. The numbers above are calculated for the case of an Einstein - de Sitter
Universe, where $w = 1 - (1 + z)^{-1/2}$.
The estimated velocity dispersion of $1150 $km s$^{-1}$ for \objectname{WL 1312.5+7252} (see
Table~\ref{tab:clumpdata}) comes from a SIS-model fit to the tangential weak shear averaged
in radial bins centered on the mass peak. This high value and the high aperture mass value
$M_{\rm ap}$ provides evidence for a
relatively low arc redshift.  The mass-to-light ratio listed in Table~\ref{tab:clumpdata} is
measured within $R_{\rm ap}$ and is based on galaxies in the color range $2.0 < V-I < 2.8$.
The lower limit of this color interval was chosen to avoid foreground galaxies associated with
\objectname{A1705}, and the upper limit was chosen to avoid background galaxies that are
significantly redder than the red sequence of \objectname{WL 1312.5+7252}.
The $M/L_B$ value is consistent with a normal luminous cluster, and assuming a redshift-dependence
$L_B \propto (z+1)$ for the
luminosity evolution, the present-day value would be $M/L_B = (268\pm 84) h$.

The only pointed X-ray observation of \objectname{WL 1312.5+7252} is
a 3.8 ksec EINSTEIN IPC exposure. In the IPC image there is diffuse emission
from the main A1705 cluster extending in the direction of \objectname{WL 1312.5+7252}.
Also, there is a hint of diffuse emission extending towards the mass structure
southeast of the main cluster. The bright X-ray point source,
\objectname{RX J1313.5+7259}, located northeast of the main cluster masks any signs of diffuse
emission that might be associated with the feature in the mass map in this direction.

\ifthenelse{\equal{\version}{_apj}}
{}{
\begin{inlinefigure}
\centering\epsfig{file=f8.eps,width=\figwidth}
\caption[Mass map of A1722.]
{The reconstructed projected mass density in the field of \objectname{A1722}. The mass map is based on the KS93 inversion method, with a Gaussian smoothing scale of $47\arcsec$.  The scale bar on the right runs from zero to the peak value of $\kappa$. The plotted contours start at zero and are spaced at intervals of $\Delta \kappa = 0.017$.}
\label{fig:a1722_massmap}
\end{inlinefigure}
}

If \objectname{WL 1312.5+7252} is at a redshift of $z=0.55$,
within a radius of $r_{500}$ it is at least a
factor three X-ray under-luminous relative to the expectation from the
mass-X-ray luminosity relation of Reiprich and B\"{o}hringer (2002).
Again, given the scatter in the mass-luminosity relation of an order of
magnitude, \objectname{WL 1312.5+7252} could be a rather normal cluster by
X-ray standards.

In conclusion, information from galaxy photometry, (probable) strong
lensing and
weak lensing all indicate the presence of a new massive, high-redshift cluster in the field
of \objectname{A1705}. Although the cluster is not detected in
existing X-ray data, the X-ray constraints on the cluster are consistent with the expected
cluster X-ray luminosity, but places the cluster at the faint end of the observed distribution
of X-ray luminosities for the given mass.
It would of course be of great interest to accurately
measure the redshift of this cluster, and of the arc, to obtain definite confirmation of the
lensed nature of the arc and to increase the accuracy of mass
estimates of \objectname{WL 1312.5+7252} from strong and weak lensing.
Furthermore, Chandra or XMM-Newton observations are crucial to
determine the hot gas content of the cluster and to independently map
the mass distribution.

\subsection{The A1722 Field}
\label{sec:a1722field}

The cluster center of \objectname{A1722} is dominated by a single
cD galaxy. A $17\arcsec$ long, thin blue gravitationally lensed arc is
visible $18\arcsec$ southwest of the cD.
As shown in the mass map (Figure~\ref{fig:a1722_massmap}), there is a second peak in
the mass distribution, \objectname{WL 1320.4+6959}, which is almost as significant as the
mass peak corresponding to \objectname{A1722} (its peak value is $\kappa = 0.166$).
Figure~\ref{fig:a1722_colsel} shows the distribution of red ($V-I > 2.2$) galaxies
in the field. A concentration of galaxies is seen close to the position of
\objectname{WL 1320.4+6959}, but with a larger spread in galaxy colors than the
\objectname{WL 1312.5+7252} cluster in the field of \objectname{A1705}.
There appears to be a general overdensity of red galaxies in the southern part
of the \objectname{A1722} field, and there are several galaxy concentrations along
a filamentary structure running along the southern part of the field. The same
general appearance is evident in the panels corresponding to $V-I > 2.0$ in
Figure~\ref{fig:a1722_colnumdens}.

The apparent early-type galaxy sequence associated with the galaxy concentration around
\objectname{WL 1320.4+6959} is less well-defined
than for \objectname{WL 1312.5+7252}, but there appears to be a concentration of galaxies
with colors around $V-I = 2.25$, corresponding to an estimated redshift of $z = 0.45\pm 0.05$.
A corresponding strong density peak is seen for galaxies in the range $1.9 < V-I < 2.5$
in Figure~\ref{fig:a1722_colnumdens}. As shown in Figure~\ref{fig:a1722darkmass},
there is an offset of $\sim 1\arcmin$ between this galaxy density peak and the

\ifthenelse{\equal{\version}{_apj}}
{}{
\begin{inlinefigure}
\centering\epsfig{file=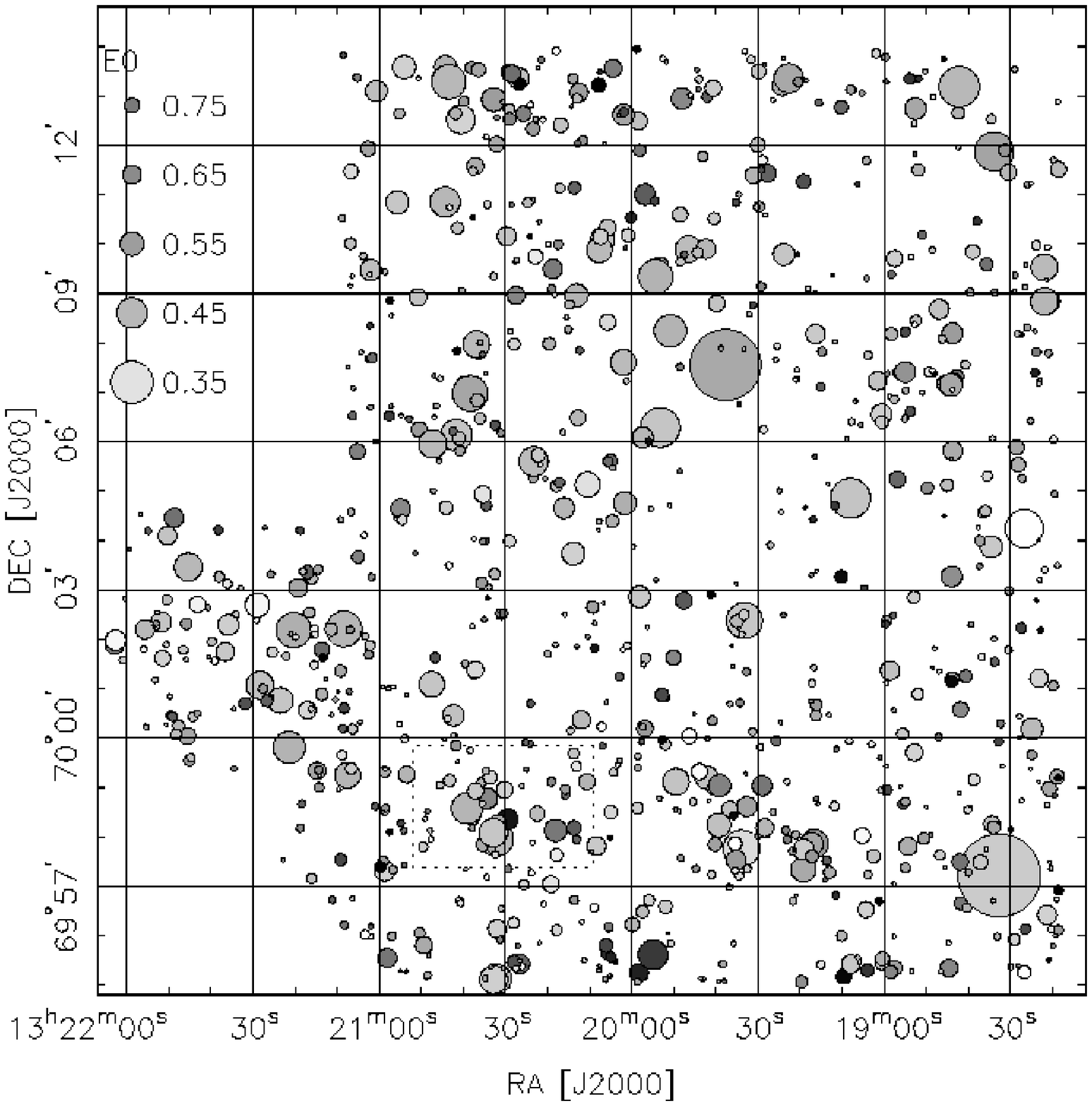,width=8cm, height=11cm}
\caption[Red galaxies in the field of A1722.]
{The reddest galaxies ($V-I > 2.2$) in the field of Abell 1722.
 The area of the circles is proportional to the $I$-band flux. The legend in the upper left corner indicates the
flux and color (assuming no evolution of the SED) of an $L_{\ast}$ E0 galaxy at a given redshift. The rectangular region denoted by the dotted lines indicates the region shown in Fig.~\ref{fig:a1722_colnumdens}. The lack of data in the lower left (southeastern) corner of the image is caused by obscuration by the guide probe when obtaining the $I$-band data for this field.}
\label{fig:a1722_colsel}
\end{inlinefigure}
}

\objectname{WL 1312.5+7252} mass peak. A red galaxy with an extended
envelope is located within the innermost contour of the galaxy density peak,
and appears to be the central galaxy of the cluster.
The two last panels of Figure~\ref{fig:a1722_colnumdens} show a moderate number density
peak of very red ($2.65 < V-I < 3.1$) galaxies located even closer to the
\objectname{WL 1320.4+6959} mass peak. A color of $V-I = 2.9$ places the galaxies
in this peak at $z \sim 1$ or higher, but any cluster at this redshift would have to
be unrealistically massive to produce the observed weak lensing signal, since most of the
source galaxies used in the lensing analysis are situated at lower redshifts.
We therefore believe that the clustering of galaxies at $z \sim 0.45$ is chiefly
responsible for producing the peak in projected mass density distribution.
The first five panels in Figure~\ref{fig:a1722_colnumdens} ($1.15 < V-I < 2.05$) also
show moderate galaxy density peaks close to \objectname{WL 1320.4+6959}. Thus, it is possible
that a projection of multiple structures along the line of sight, some of them possibly associated
with \objectname{A1722}, are contributing to
the \objectname{WL 1320.4+6959} mass peak. Alternatively, we may just be seeing galaxies of
different colors and morphological types at a single redshift. In this case, the significant scatter
seen in the red cluster sequence suggests that the cluster is less dense and has a larger
fraction of late-type galaxies than \objectname{WL 1312.5+7252}, which would be expected
for a cluster which is still forming. Spectroscopic determinations of galaxy redshifts in the area
of \objectname{WL 1320.4+6959} would be required to conclusively distinguish between these
two scenarios.

From an ASCA SIS 66 ksec exposure, the upper limit on the X-ray luminosity from
\objectname{WL 1320.4+6959} is a factor two below the expectation from
the mass-luminosity relation of Reiprich \& B\"{o}hringer (2002). However,
the upper limit on the X-ray luminosity is uncertain by a factor two
due to the uncertain subtraction of X-ray emission from the main cluster
and a bright point source (seen in an archival 27 ksec ROSAT HRI exposure)
southwest of the main cluster.
Although \objectname{WL 1320.4+6959} is
X-ray underluminous, existing X-ray data are consistent with a
cluster at redshift of $z \sim 0.45$.

The mass-to-light ratio listed in Table~\ref{tab:clumpdata} is
measured within $R_{\rm ap}$ and is based on galaxies in the color range $2.0 < V-I < 2.6$.
The lower limit of this color interval was chosen to avoid foreground galaxies associated
with \objectname{A1722} (see Figure~\ref{fig:a1722_colnumdens}), and the upper limit was
chosen to avoid background galaxies that are significantly redder than the apparent red
sequence associated with \objectname{WL 1320.4+6959}.
The $M/L_B$ value is consistent with a normal luminous cluster, and assuming $L_B \propto (z+1)$
luminosity evolution, the present-day value would be $M/L_B = (429\pm 171) h$.

\ifthenelse{\equal{\version}{_apj}}
{}{
\begin{figure*}
\centering\epsfig{file=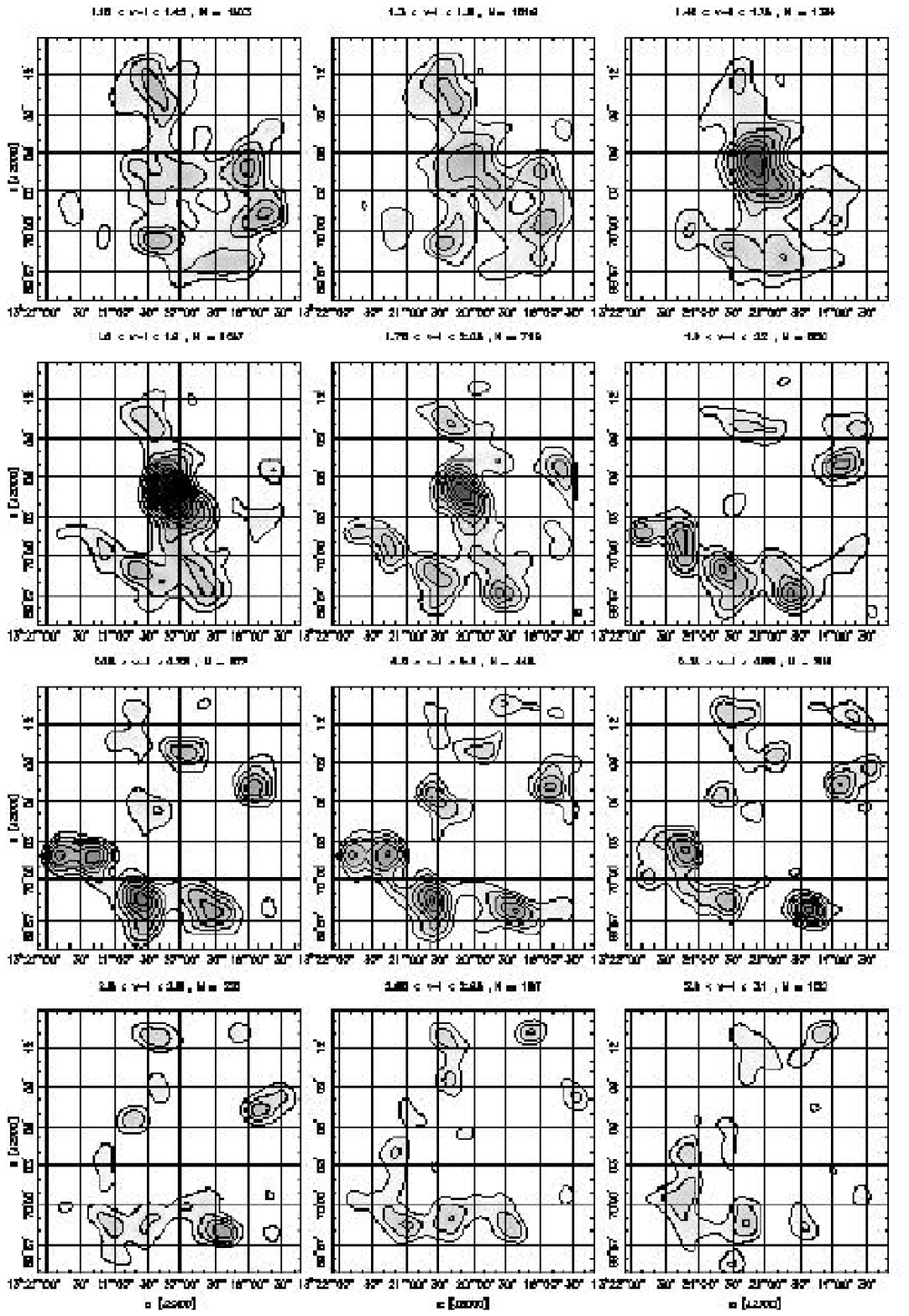,width=15cm}
\caption[Color-selected galaxy density distribution around A1722.]
{The surface number density distribution of $I < 23.0$ galaxies in various $V-I$ color intervals in the field of \objectname{A1722}. See the caption of Fig.~\ref{fig:a959_colnumdens} for details. }
\label{fig:a1722_colnumdens}
\end{figure*}
}

\section{Discussion}
Using weak gravitational lensing data, we have in the previous paragraphs identified three
prominent (projected) mass concentrations representing new galaxy cluster (or sub-cluster) candidates.
We use two-color, $V$- and $I$-band photometry to investigate the nature of these structures
and make rough redshift estimates.

The first mass concentration, \objectname{WL 1017.3+5931}, is the most enigmatic of these objects.
The morphology of the mass peak and the associated X-ray emission
suggest a possible association with the nearby $z = 0.29$ cluster
\objectname{A959}, but we find no overdensity of early-type cluster galaxies at the
location of \objectname{WL 1017.3+5931}. There is also no strong evidence for clustering
of galaxies at any other redshift at this position. This object remains a good candidate
for an optically ``dark'' (sub-)cluster. The upper limit on the X-ray
luminosity leaves the possibility that it is an X-ray (underluminous)
cluster at or beyond the redshift of \objectname{A959}. Deep X-ray
data of this system woulde be particularly interesting in order to
accurately measure its hot gas content.

The second mass concentration, \objectname{WL 1312.5+7252}, appears to
be associated with a rich cluster at $z \sim 0.55$ which also acts as a strong lens.
This structure is associated with a prominent overdensity of red galaxies and has a
prominent red galaxy sequence. Thus, it is likely to constitute a single strong physical
overdensity of dark matter and galaxies,
rather than being caused by a line-of-sight projection of lesser structures.
Its $M/L_B$ value is similar to the typical values of clusters
selected from baryonic tracers. The conservative upper limit on the X-ray
luminosity of \objectname{WL 1312.5+7252} is consistent with a cluster
at $z \sim 0.55$ with the mass determined from weak lensing. However,
\objectname{WL 1312.5+7252} could well be an X-ray dark mass
concentration, containing only a small amount of hot gas.

The third mass concentration, \objectname{WL 1320.4+6959}, is associated with an
overdensity of galaxies at an estimated redshift of $z \sim 0.45$, but these galaxies show a
larger spread in $V-I$ color than the galaxies in the \objectname{WL 1312.5+7252} cluster.
However, the upper limit on its X-ray emission does not allow us to
rule out that \objectname{WL 1320.4+6959} is a single cluster at
$z\sim 0.45$ which is X-ray underluminous and possibly X-ray dark. It is thus unclear
whether this density peak represents a chance superposition of objects at different
redshifts or whether it represents a single cluster at $z \sim 0.45$ which is still
forming and is not yet virialized. Weinberg \& Kamionkowski (2002) estimate
the abundances of non-virialized, X-ray underluminous protoclusters that will be detectable
in weak lensing surveys, and they find that such systems are likely to be detected in surveys
comparable to ours. The $M/L_B$ value we find is typical for a normal
optically luminous cluster.

At this point we may speculate whether it is significant that we find three mass
concentrations with $\kappa > 0.15$
based on 1.0 deg$^2$ of imaging of fields containing massive clusters,
while Wilson et al.\ (2001) find no such objects from similar data (1.5 deg$^2$ of imaging) of
blank fields. We also recall that the two optically dark mass concentrations previously discovered by
Erben et al.\ (2000) and Umetsu \& Futamase (2000) were both found in the fields of massive
clusters.
In the case of \objectname{WL 1312.5+7252} we are seeing a mass concentration which
is clearly physically unrelated to the nearby Abell cluster. For \objectname{WL 1320.4+6959}
it is hard to draw an equally firm conclusion, since, as noted in
\S~\ref{sec:a1722field}, there may be a significant contribution
to the projected mass density from structures at the redshift of \objectname{A1722}.

\ifthenelse{\equal{\version}{_apj}}
{}{
\begin{figure*}
\centering\epsfig{file=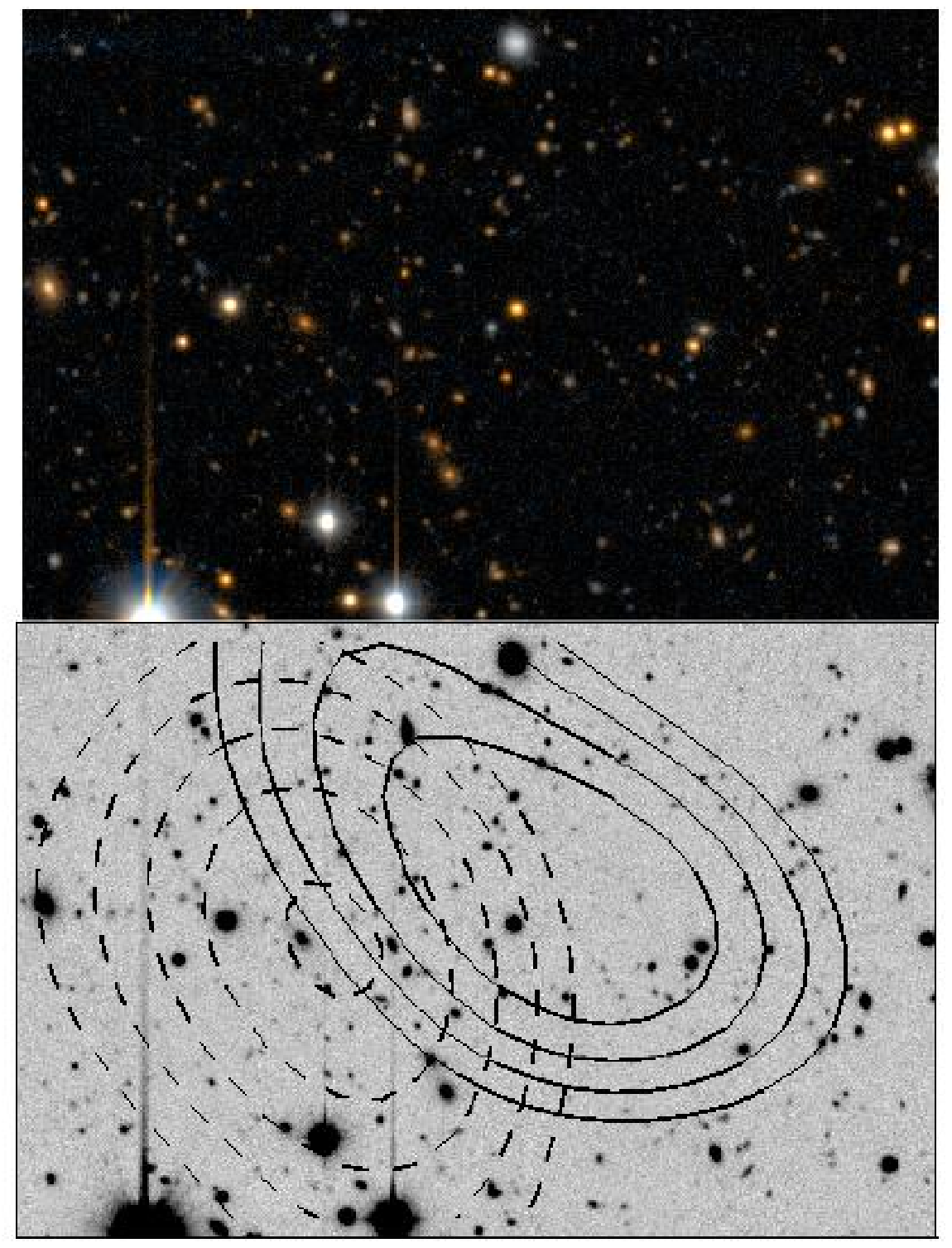}
\caption[Area around secondary mass peak in A1722.]
{The plots show a $225\arcsec \times 150\arcsec$ region around the \objectname{WL 1320.4+6959} mass peak in the field of \objectname{A1722}. Top: ``True color'' image based on 4.5h of integration in the $V$-band and 4h of integration in the $I$-band. Bottom: Solid lines show contours of the projected mass density $\kappa$ indicating the \objectname{WL 1320.4+6959} mass peak. Contour levels start at $\kappa = 0.1$ and are plotted at intervals of $0.02$ in $\kappa$. The dashed lines are contours of the smoothed galaxy density distribution in the $2.2 < V-I < 2.5$ subpanel in Figure~\ref{fig:a1722_colnumdens}.}
\label{fig:a1722darkmass}
\end{figure*}
}

The photometric data for \objectname{WL 1017.3+5931} do not provide any strong evidence for
an association with nearby \objectname{A959}, but the morphology of the mass peak does suggest a
possible physical link between the two objects. In a recent study, White, van Waerbeke \& Mackey (2002)
show that significant peaks in the projected density distribution, resembling clusters with virial
masses of $10^{14} - 3\times 10^{14} h^{-1} M_{\odot}$ may be generated by line-of-sight projections
of multiple correlated structures with $M < 10^{14} h^{-1} M_{\odot}$. It is possible that
\objectname{WL 1017.3+5931} is such an object, generated by a superposition of
$\sim 10^{13} h^{-1} M_{\odot}$ halos within the overdensity associated with \objectname{A959}.
This scenario would naturally explain the low X-ray luminosity, but does not predict an unusually high
$M/L_B$ value or the lack of an associated peak in the galaxy density distribution.

Clearer answers to the nature of weak lensing-detected mass concentrations may come soon
from systematic cluster searches in the deep wide-field imaging data sets currently used for
measurements of ``cosmic shear''. The sky area collectively probed by such surveys to similar depth
is now at least an order of magnitude larger than the sky area we study here
(see e.g., van Waerbeke et al.\ 2001).

Our results also demonstrate the usefulness of multi-color photometry and color slicing techniques
when interpreting results from weak lensing cluster searches. At redshifts $z \sim 0.5$ and higher,
even rich clusters do not represent strong galaxy density enhancements in single-passband imaging data.
In the case of \objectname{WL 1320.4+6959}, an optical counterpart to the mass concentration
could not be identified before data in a second passband had been obtained.
It is also clear that X-ray data from either Chandra or XMM-Newton are required in order to detect or
tightly constrain the hot gas content of the detected (sub-)clusters.

The methodology for cluster searches may be further refined in the future by developing
a more objective and quantitative search algorithm that combines weak gravitational
lensing information with e.g, the cluster-red-sequence method of Gladders \& Yee and/or X-ray data.
The X-ray data would be particularly useful for separating line-of-sight superpositions of less massive
objects (which would be a significant source of noise and bias for both optical and weak lensing data;
see e.g., Hoekstra 2001; White et al. 2002) from genuine deep potential wells.
This would greatly improve the power of such surveys to constrain cosmological models.

\acknowledgements

We thank the anonymous referee for suggestions that have
improved our work and its presentation. We also thank
Harald Ebeling, Jens Hjorth, Henk Hoekstra, Gerry Luppino, Somak Raychaudhury, and Gillian Wilson
for useful comments and discussions.
We thank the staff of the University of Hawaii 2.24m telescope and the Nordic
Optical Telescope for support during our observing runs.
HD gratefully acknowledges support from a doctoral research fellowship awarded by the
Research Council of Norway. HD and PBL thank the Research Council
of Norway for travel support. KP acknowledges support from the Danish National Research Council.
This research has made use of the NASA/IPAC Extragalactic Database (NED) which
is operated by the Jet Propulsion Laboratory, California Institute of
Technology, under contract with the National Aeronautics and Space Administration.
This research has made use of data obtained from the High Energy
Astrophysics Science Archive Research Center (HEASARC), provided by
NASA's Goddard Space Flight Center.


\begin{thebibliography}{}


\bibitem[Abazajian, Fuller, \& Tucker(2001)]{2001ApJ...562..593A} 
Abazajian, K., Fuller, G.~M., \& Tucker, W.~H.\ 2001, \apj, 562, 593. 

\bibitem[Abell (1958)]{ab}  Abell, G. O. 1958, \apjs, 3, 211

\bibitem[Bertin \& Arnouts (1996)]{1996A&AS...447L..81B} Bertin, E., \& Arnouts, S. 1996, \aaps, 117, 393

\bibitem[Briel \& Henry]{bh} Briel, U., \& Henry, J. P. 1993, \aap, 278, 379

\bibitem[B{\" o}hringer et al.(2000)]{2000ApJS..129..435B} B{\" o}hringer, 
H.~et al.\ 2000, \apjs, 129, 435 

\bibitem[Carlberg, Yee, \& Ellingson(1997)]{1997ApJ...478..462C} Carlberg, R.~G., Yee, H.~K.~C., \& 
Ellingson, E.\ 1997, \apj, 478, 462 

\bibitem[Dahle et al.\  2002]{da} Dahle, H., Kaiser, N., Irgens, R. J., Lilje, P. B., \& Maddox, S. J. 2002, 
\apjs, 139, 313 (Paper I)

\bibitem[Dickey and Lockman 1990]{dl} Dickey, J.M. \& Lockman,
F.J.,1990, Ann.R.A\& A., 28, 215

\bibitem[Erben et al.(2000)]{2000A&A...355...23E} Erben, T., van Waerbeke, 
L., Mellier, Y., Schneider, P., Cuillandre, J.-C., Castander, F.~J., \& 
Dantel-Fort, M.\ 2000, \aap, 355, 23 

\bibitem[Ettori \& Fabian(1999)]{1999MNRAS.305..834E} Ettori, S.~\& Fabian, 
A.~C.\ 1999, \mnras, 305, 834 

\bibitem[Fahlman et al.\ 1994]{fksw} Fahlman, G., Kaiser, N., Squires, G., \& Woods, D. 1994, \apj, 437, 56

\bibitem[Finoguenov et al.\ 2001]{frb} Finoguenov, A.; Reiprich, T. H.; Böhringer, H.\ 2001, \aap, 368, 749

\bibitem[Fischer(1999)]{1999AJ....117.2024F} Fischer, P.\ 1999, \aj, 117, 2024 

\bibitem[Gladders \& Yee(2000)]{2000AJ....120.2148G} Gladders, M.~D.~\& 
Yee, H.~K.~C.\ 2000, \aj, 120, 2148 

\bibitem[Gray et al.(2001)]{2001MNRAS.325..111G} Gray, M.~E., Ellis, R.~S., 
Lewis, J.~R., McMahon, R.~G., \& Firth, A.~E.\ 2001, \mnras, 325, 111 

\bibitem[Hansen, Lesgourgues, Pastor, \& Silk(2002)]{2002MNRAS.333..544H} 
Hansen, S.~H., Lesgourgues, J., Pastor, S., \& Silk, J.\ 2002, \mnras, 333, 
544 

\bibitem[Hoekstra(2001)]{2001A&A...370..743H} Hoekstra, H.\ 2001, \aap, 
370, 743 

\bibitem[Hoekstra, Franx, Kuijken, \& van 
Dokkum(2002)]{2002MNRAS.333..911H} Hoekstra, H., Franx, M., Kuijken, K., \& 
van Dokkum, P.~G.\ 2002, \mnras, 333, 911 

\bibitem[Irgens, Lilje, Dahle, \& Maddox(2002)]{2002ApJ...579..227I} 
Irgens, R.~J., Lilje, P.~B., Dahle, H., \& Maddox, S.~J.\ 2002, \apj, 579, 227 (Paper II)

\bibitem[{{Kaiser}(2000)}]{kaiser00}
{Kaiser}, N. 2000, \apj , 537, 555

\bibitem[Kaiser \& Squires (1993)]{ks} Kaiser, N. \& Squires,G. 1993, \apj, 404, 441, KS93

\bibitem[Kaiser et al.\ 1994]{ksfw} Kaiser, N., Squires, G.,
Fahlman, G., \& Woods, D. 1994, in ``Clusters of Galaxies'', XXIXth
Rencontres de Moriond, ed. F. Durret, A.  Mazure, and J. Tran Thanh Van (Gif-sur-Yvette: Edition Fronti\`eres) 

\bibitem[{{Kaiser} {et~al.}(1999){Kaiser}, {Wilson}, {Luppino}, \&
  {Dahle}}]{kwld99}{Kaiser}, N., {Wilson}, G., {Luppino}, G., \& {Dahle}, H. 1999, preprint (astro-ph/9907229)

\bibitem[Lubin(1996)]{1996AJ....112...23L} Lubin, L.~M.\ 1996, \aj, 112, 23 

\bibitem[Miralles et al.(2002)]{2002A&A...388...68M} Miralles, J.-M.~et al.\ 2002, \aap, 388, 68 

\bibitem[Mushotzky \& Scharf(1997)]{1997ApJ...482L..13M} Mushotzky, 
R.~F.~\& Scharf, C.~A.\ 1997, \apjl, 482, L13 

\bibitem[Reiprich & B\"{o}hringer (2002)]{rb} Reiprich, T. H. \& B{\"o}hringer, H. 2002, \apj, 567, 716 

\bibitem[Smail, Edge, Ellis, \& Blandford(1998)]{1998MNRAS.293..124S} 
Smail, I., Edge, A.~C., Ellis, R.~S., \& Blandford, R.~D.\ 1998, \mnras, 293, 124 

\bibitem[Stanford, Eisenhardt, \& Dickinson(1998)]{1998ApJ...492..461S} 
Stanford, S.~A., Eisenhardt, P.~R., \& Dickinson, M.\ 1998, \apj, 492, 461 

\bibitem[Umetsu \& Futamase(2000)]{2000ApJ...539L...5U} Umetsu, K.~\& 
Futamase, T.\ 2000, \apjl, 539, L5 

\bibitem[Van Waerbeke et al.(2001)]{2001A&A...374..757V} Van Waerbeke, L.~et al.\ 2001, \aap, 374, 757. 

\bibitem[Weinberg \& Kamionkowski(2002)]{2002MNRAS.337.1269W} Weinberg, 
N.~N.~\& Kamionkowski, M.\ 2002, \mnras, 337, 1269 

\bibitem[White, van Waerbeke, \& Mackey(2002)]{2002ApJ...575..640W} White, 
M., van Waerbeke, L., \& Mackey, J.\ 2002, \apj, 575, 640 

\bibitem[Wilson, Kaiser, \& Luppino(2001)]{2001ApJ...556..601W} Wilson, G., 
Kaiser, N., \& Luppino, G.~A.\ 2001, \apj, 556, 601 

\bibitem[Wittman et al.(2001)]{2001ApJ...557L..89W} Wittman, D., Tyson, 
J.~A., Margoniner, V.~E., Cohen, J.~G., \& Dell'Antonio, I.~P.\ 2001, 
\apjl, 557, L89 


\end{thebibliography}
\end{document}